\documentclass[conference,letterpaper,twocolumn,10pt]{IEEEtran}

\usepackage{paralist}
\usepackage{needspace}
\newcommand{\para}[1]{\Needspace{3\baselineskip}\noindent\textbf{#1. }\ignorespaces}
\usepackage{graphicx}
\usepackage{enumitem}    
\usepackage[normalem]{ulem}
\usepackage{amsmath}
\usepackage{amssymb}
\usepackage{xspace}
\usepackage{tabularx}      
\usepackage{textcomp}      
\usepackage[table,dvipsnames,svgnames]{xcolor}
\usepackage{adjustbox}
\usepackage{stfloats}
\usepackage{cite}

\PassOptionsToPackage{hyphens}{url}\usepackage[hidelinks]{hyperref}
\hypersetup{
    colorlinks=true,
    linkcolor=ForestGreen,
    citecolor=ForestGreen,
    urlcolor=RoyalBlue,
    filecolor=IndianRed,
    pdftitle={Performance Verification of the AmpereOne CPU Core},
    pdfauthor={Doa'a Al-Otoom, Nick Kelly, Mahesh Madhav}, 
    pdfsubject={Benchmarks}, 
    pdfkeywords={Standardization, Benchmarks, CPU Performance}, 
    pdfnewwindow=true,
    pdfdisplaydoctitle=true,
    bookmarksopen=true
}

\begin{document}
\pagestyle{plain} 
\newcommand{\Red}[1]{{\color{red} #1}}
\newcommand{\ignore}[1]{}
\newcommand{\asm}[1]{\texttt{#1}}
\newcommand{\sys}[1]{\texttt{#1}}
\newcommand{\kw}[1]{\textit{#1}}
\newcommand{\kwb}[1]{\textbf{#1}}
\newcommand{\type}[1]{\textit{#1}}
\newcommand{\Response}[1]{{\color{black} #1}}
\newcommand{\XXX}[1]{\Red{\textbf{XXX[}#1\textbf{]}}}
\newcommand{\tocite}[1]{\Red{CITE:\cite{#1}}}
\newcommand{\toref}[1]{\Red{REF:\ref{#1}}}
\newcommand{\parasub}[1]{\smallskip\noindent\textit{{#1:}\xspace}}
\newcommand{\mahesh}[1]{\textcolor{purple}{#1}}
\newcommand{\anyone}[1]{\textcolor{blue}{#1}}
\newcommand{\niparagraph}[1]{\noindent\textbf{\textsf{#1}\hspace{0.5em}}}
\newcommand\TODO[1]{\textcolor{red}{TODO: #1}}
\newcommand\todo[1]{\textcolor{red}{TODO: #1}}

\newenvironment{CompactItemize}%
  {\begin{list}{$\blacktriangleright$}%
    {\leftmargin=\parindent \itemsep=2pt \topsep=2pt
     \parsep=0pt \partopsep=0pt}}%
  {\end{list}}
\renewcommand{\labelitemi}{$\blacktriangleright$}

\newcommand{\malloc}{{\texttt{Malloc}}}
\newcommand{\linklist}{{\textsf{Linked-List}}}

\definecolor{train}{RGB}{43, 77, 137}
\definecolor{trigger_bop}{RGB}{0, 122, 55}
\definecolor{trigger_sbop}{RGB}{164, 0, 0}
\definecolor{trigger_nl}{RGB}{165, 76, 15}
\definecolor{trigger_sp}{RGB}{158, 120, 0}
\definecolor{overflow}{RGB}{202, 2, 102}
\definecolor{launch}{RGB}{112, 48, 160}

\definecolor{cond}{RGB}{27, 117, 177}
\definecolor{call_ind}{RGB}{215, 37, 37}
\definecolor{uncond_ind}{RGB}{39, 157, 39}
\definecolor{return}{RGB}{146, 99, 187}
\definecolor{miss_event}{RGB}{255, 80, 120}
\definecolor{alloc_event}{RGB}{255, 119, 39}
\definecolor{hit_event}{RGB}{0, 160, 255}
\definecolor{dealloc_event}{RGB}{120, 120, 255}

\DeclareRobustCommand{\misshighlightnarrow}[1]{%
  {%
    \setlength{\fboxsep}{0pt}
    \colorbox{miss_event}{%
      \strut\textcolor{white}{{\sffamily\bfseries #1}}%
    }%
  }%
}

\DeclareRobustCommand{\allochighlightnarrow}[1]{%
  {%
    \setlength{\fboxsep}{0pt}%
    \colorbox{alloc_event}{%
      \strut\textcolor{white}{{\sffamily\bfseries #1}}%
    }%
  }%
}

\DeclareRobustCommand{\hithighlightnarrow}[1]{%
  {%
    \setlength{\fboxsep}{0pt}%
    \colorbox{hit_event}{%
      \strut\textcolor{white}{{\sffamily\bfseries #1}}%
    }%
  }%
}

\DeclareRobustCommand{\deallochighlightnarrow}[1]{%
  {%
    \setlength{\fboxsep}{0pt}%
    \colorbox{dealloc_event}{%
      \strut\textcolor{white}{{\sffamily\bfseries #1}}%
    }%
  }%
}

\title{{Performance Verification of the AmpereOne\textsuperscript{\textregistered} CPU Core}}

\author{%
\IEEEauthorblockN{Doa'a Al-Otoom, Nicholas Kelly, Aaron Lindsay, Agreen Ahmadi, Ananth Kumar,}
\IEEEauthorblockN{ Anthony Faubert, Benjamin Chaffin, Bret Toll, Eric Wallace, Lucas Crowthers,}
\IEEEauthorblockN{Mark Charney, Michael Chin, Michael Spradling, Scott Witscher, Sriyash Caculo,}
\IEEEauthorblockN{Syed Fakhri, Vivek Kothapalli, William Freelove, Mahesh Madhav}
\IEEEauthorblockA{Ampere Computing -- Portland, OR, and Raleigh, NC, and Santa Clara, CA, USA}
}


\maketitle
\begin{abstract}

As process technology scaling slows, microarchitectural innovation has become the primary driver of performance gains, making pre-silicon Performance Verification (PV) more critical than ever. This paper presents the industrial-scale PV methodology applied across four generations of the AmpereOne\textsuperscript{\textregistered} custom CPU core, centered on the cycle-accurate correlation of the RTL design against a trace-driven performance model. The methodology integrates data-driven workload curation, a high-frequency daily regression system, and a unified event-stream framework for analysis. We demonstrate this methodology through case studies of the Branch Prediction Unit and L2 Prefetcher, highlighting a hierarchical strategy that first isolates individual units for focused correlation before proceeding to full-core verification. The results demonstrate that this disciplined, iterative process is indispensable for avoiding costly post-silicon bugs and ensuring complex processors meet their performance targets. We end with a look towards the future of PV in the microprocessor industry.
\end{abstract}

\section{Introduction}

The verification of a new microprocessor represents the single largest investment in its design cycle. This effort is broadly partitioned into two fundamental disciplines: functional and performance verification. Functional Design Verification (DV), concerned with logical correctness, has received significant academic attention \cite{testge_for_uproc, test_plan, archval} and benefits from the ability to compare against a known correct output; but Performance Verification (PV) presents a more ambiguous challenge. It lacks a definitive ``oracle'' as accurately predicting the performance of a given workload on a complex new microarchitecture is notoriously difficult. Nevertheless, identifying and eliminating performance bugs has become more critical than ever. As the cadence of process technology scaling slows, the primary source of generational performance and efficiency gains shifts to microarchitectural innovation, and performance bugs can erode these hard-won improvements.

The push for comprehensive pre-silicon PV is underscored by the high costs of post-silicon bug discovery. Performance discrepancies found after tape-out can necessitate a metal layer or even full-chip resynthesis, which are quite costly not just monetarily but also for schedule delays and missed market opportunities. Even bugs that do not require a respin often demand expensive mitigations which cap performance and weaken the product's competitive position. Because performance anomalies tend to be deeply embedded in the microarchitecture, they are among the most difficult class of bugs to resolve post-silicon, often lacking a clean software-level fix. 

Thus, pre-silicon performance evaluation and verification has been a critical component during the design and development of high-end microprocessors for decades \cite{powerpc_1994, perf_val_1999, powerpc_perf, sparcv9_perf, intel, powerpc_pv, freescale_pv, power10_pv, pv_china}. CPU architects construct software models of the CPU to simulate hardware behaviors. In an industrial setting, these models serve four primary purposes: (1) a sandbox for architectural pathfinding, (2) a reference for PV against the RTL design, (3) a forecasting tool for product planning, and (4) a pre-silicon platform for early software development. This paper focuses on the second purpose: pre-silicon performance verification, and the process of exposing and fixing bugs in the RTL of the AmpereOne\textsuperscript{\textregistered} core.



This model-based correlation comes with its own challenges. The industry has long acknowledged that building and maintaining a sufficiently accurate performance model is a time-consuming and labor-intensive task. Prior work has shown that performance estimates from simulation can diverge significantly from real hardware \cite{sim_trace_driven, sim_error}, and the process of diagnosing discrepancies often devolves into tedious manual analysis. On top of this, as these models are often developed in parallel with the RTL, they are susceptible to common-mode errors where the same bug may exist in both the reference and the design, rendering the issue undetectable \cite{archval}. Despite these inherent difficulties, the rigorous comparison of RTL execution against a reference model remains the industry's most effective methodology for ensuring a design is on track to meet its performance goals.

This paper presents a detailed case study of the pre-silicon performance verification methodology applied to the Ampere custom CPU core IP. We describe the infrastructure, workload selection, and analytical processes used to correlate the RTL design against its microarchitectural performance model. We dive deep into a couple of key features to show how we accomplished correlation at the levels of both the full core and individual units. By detailing our approach to identifying, diagnosing, and resolving performance bugs prior to tape-out, we provide insight into a practical application of performance verification and demonstrate its role in delivering a competitive, high-performance product. 

\Response{The novelty of this work is not the claim that RTL should be compared against a model; rather it is the industrial methodology for making that comparison actionable on modern CPU microarchitectures. The key mechanisms we describe are shared model/RTL event semantics, trace-driven wrong-path compensation, unit-level replay for timing-sensitive structures, and daily outlier triage tied to domain expertise. To that end, we focus here on single-core microarchitectural PV; multicore and SoC-level performance issues, such as shared-data contention and coherency effects, were verified with different methods and are left to future publications.}

\section{Background}

The methodology and insights presented in this paper were developed and refined across four generations of Ampere's custom cores. Each successive generation introduced new microarchitectural features and more aggressive performance targets, which in turn demanded continuous enhancement of the performance verification flow. This iterative improvement proved useful, occasionally uncovering performance bugs that had escaped detection in previous product cycles and reinforcing the importance of an evolving, ever-more-rigorous methodology.

The end-to-end performance verification flow integrates several parallel and sequential activities that culminate in the final correlation effort. The key stages are as follows:

\para{Workload and Model Development} The process begins with the selection of representative workloads, from which execution traces are generated. These traces serve as the common stimulus for both the model and the RTL. In parallel, the model itself is developed and refined, serving as the golden reference of behavior.

\para{RTL Implementation and Instrumentation} The design team implements the core microarchitecture in SystemVerilog, guided by architectural specifications and the behavior of the performance model. In parallel, sophisticated observability tooling is developed for both the model and the RTL. This instrumentation generates detailed event outputs by tracking instructions and micro-operations through all pipeline stages. To conceptualize this, one can draw an analogy to motion-capture technology \cite{mocap}: the events are like markers on an actor, and our visualization tools create an abstract representation of the pipeline's `motion'. The verification task is then to confirm that the behaviors of the model and the RTL are identical by comparing these representations.

\para{Iterative Correlation Loop} The core of the PV process is a continuous cycle of correlation. Traces and targeted microbenchmarks are executed on both the performance model and RTL. Key performance metrics (e.g., Instructions Per Cycle, cache miss rates) are compared, and any significant deviation triggers an in-depth debugging session. These sessions rely heavily on the pipeline visualization tools to isolate the root cause of the discrepancy in either the model or the RTL. This cycle of running, comparing, and debugging is repeated daily/weekly throughout the project, driving convergence between the model and the design until a high degree of confidence in the core's performance is achieved.

The following sections will provide a more detailed examination of these processes and present specific case studies from this verification campaign.

\section{Performance Model}

The foundation of the performance verification methodology at Ampere is a custom-built performance simulator named \textit{Panthera} \cite{panthera}. The Panthera framework is implemented in C++ and leverages the SystemC library to model the core microarchitecture as well as the system-level mesh interconnect. Panthera is a trace-driven simulator, a design choice that contrasts with the execution-driven models used elsewhere in the industry. This approach was chosen to balance development effort with predictive accuracy, providing a favorable return on investment for a brand new modeling infrastructure. The efficacy of this decision is validated by strong post-silicon correlation; for instance, our pre-silicon projections for multi-tasking benchmarks like SPEC CPU2017 consistently fall within 5\% of measured hardware performance on a 192-core server \cite{ampere_mav, panem}. This paper details the verification process that enables such accuracy on AmpereOne\textsuperscript{\textregistered}.


\section{Workload Curation and Trace Collection}

\begin{figure}[b] 
    \centering
    \includegraphics[width=\columnwidth]{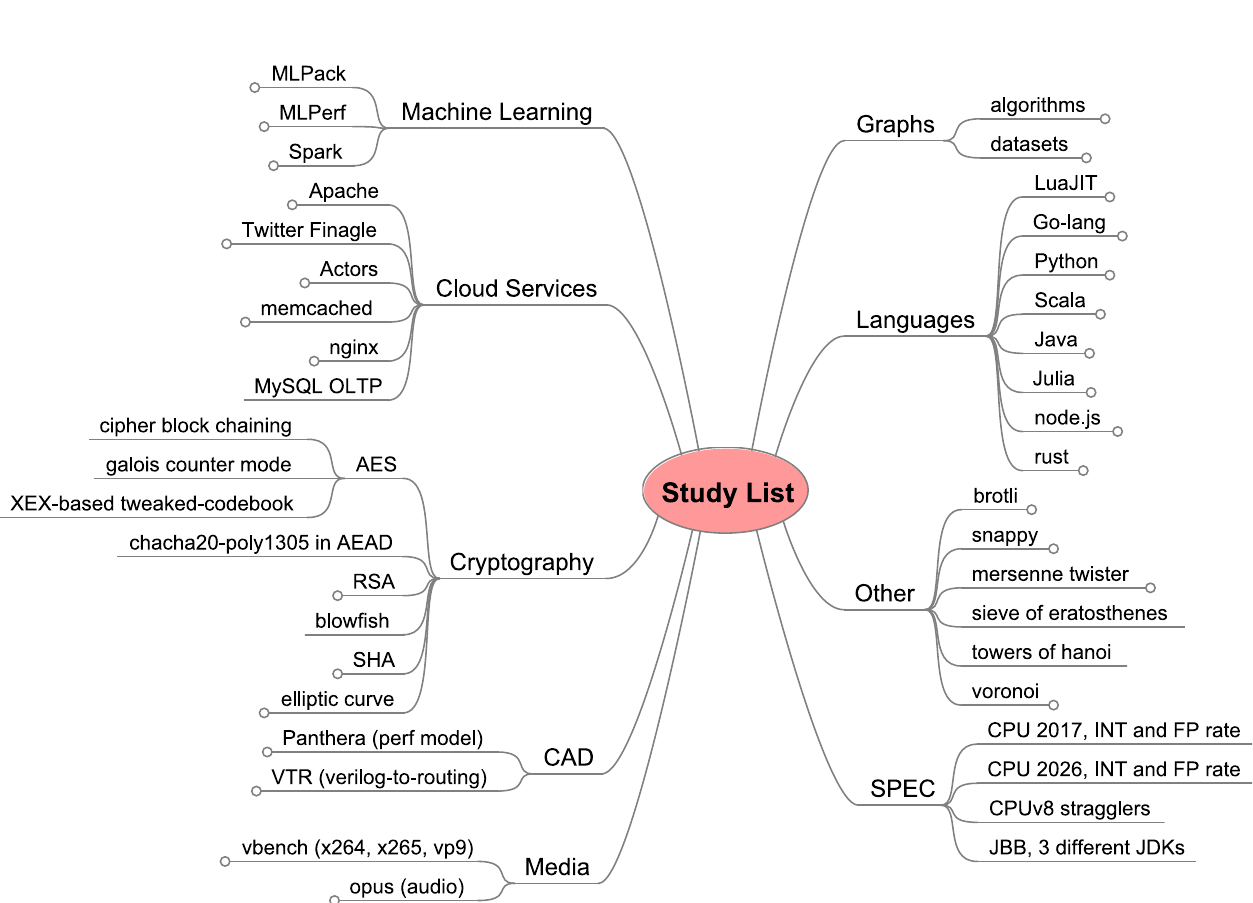}
    \caption{Sample taxonomy of workloads in the core study list \cite{study_list_2024}}
    \label{fig:study list}
\end{figure}

Effective PV hinges on a meticulously curated set of input stimuli. The foundation for our workload suite is a ``study list'' developed during the architectural pathfinding phase, which contains applications and behaviors identified as critical to Ampere's target cloud computing market \cite{mlperf, dacapo_25, ren_1, cpu2026, cpu2017, tpcc, specjbb2012, vbench}. A representative sample of this application mix is shown in \autoref{fig:study list}. These full applications are then processed to generate targeted traces suitable for simulation \cite{study_list_2024}.

Our process begins by using SimPoint, a well-established methodology that analyzes a program's execution using Basic Block Vectors (BBVs) to identify a small number of representative execution intervals \cite{simpoint}. These intervals, or ``SimPoints,'' collectively model the performance characteristics of the entire application, avoiding the need for prohibitively long simulations. The tracing infrastructure used to capture these intervals is built upon QEMU \cite{qemu_trace0, qemu_trace1}. A key advantage of using the QEMU emulator is the ability to trace execution across different Exception Levels (ELs), capturing both user-space (EL0) and operating system kernel (EL1) activity \cite{arm_exception_levels}. For cloud-native workloads, where OS interactions like system calls and memory management constitute a significant portion of execution time \cite{google_wsc}, this full-system visibility is key to identifying performance bottlenecks that would be missed by tracing only user-mode execution. The captured traces, typically on the order of 10 million instructions for rapid analysis in the Panthera model, are truncated to approximately 100,000 instructions for the more detailed and resource-intensive simulations in RTL.

While the study list ensures application and domain relevance, the primary goal of performance verification is comprehensive microarchitectural feature coverage. To achieve this, our stimulus suite is composed of two distinct but complementary categories: directed microbenchmarks and diverse application snippets.

\para{Directed Microbenchmarks} \label{directed_bm} These are small, hand-crafted programs, typically a few thousand instructions long, designed to stress a single microarchitectural capability or pipeline interaction \cite{pv_unittest_2001, cliffs}. These benchmarks form the first line of defense in our correlation flow, providing a focused and easily debuggable stimulus for initial bring-up. For example, before running complex application snippets, we first validate fundamental behaviors by running microbenchmarks that isolate L1 and L2 cache miss latencies. A mismatch in these simple tests indicates a foundational discrepancy that must be resolved before proceeding to more complex workloads. This hierarchical approach allows us to methodically stabilize the design early in the verification cycle. Once correlated, these microbenchmarks are integrated into the pre-commit regression suite, creating an automated guardrail that prevents future code submissions from regressing against the established performance baseline.

\para{Application Snippets} \label{app_snippets} To ensure maximum diversity of microarchitectural behaviors, we employ a data-driven selection process rather than random sampling. The entire repository of snippets is first run through the Panthera model to generate a feature vector for each one, composed of 25 key microarchitectural statistics. These include high-level indicators like instructions per cycle (IPC) as well as more granular data such as execution pipe occupancy, resource stall counts, and prefetcher efficacy. We then apply k-means clustering to this data, which groups snippets with similar hardware behaviors. This technique is similar to contemporary research that optimizes pre-silicon test selection through unsupervised machine learning \cite{presil_workloads_ml}.
The clusters allow us to curate distinct sets of tests for different purposes. A compact set of approximately 500 snippets, each representing a unique behavioral cluster, is assembled into an RTL ``perf-check-500'' regression suite \cite{perf_check_500}. This provides logic designers with rapid, broad-spectrum feedback on their new features. For more exhaustive analyses, wider sets of 2,000-cluster snippets are selected for deep-dive investigations by the PV team to correlate the design and analyze performance outliers.

To ensure high-fidelity analysis that accurately reflects execution on silicon, we pre-populate the state of key microarchitectural structures before simulation. Each snippet package contains not only the detailed instruction trace for the region of interest, but also a data stream of memory accesses and branches from billions of instructions of prior execution. This allows the simulation to begin with the caches and the Branch Prediction Unit (BPU) in a pre-conditioned or warm state. We decided to exclude the Translation Lookaside Buffer (TLB) from this warm-up methodology based on prior industry experience indicating a low return on investment. The complexity of accurately reconstructing the full TLB state, including all permission and status bits, is exceptionally high and prone to infrastructure-level bugs. Also, since TLBs typically converge to a warm state rapidly during the initial phase of snippet execution, and get frequently blown away through page invalidations, the impact on overall performance fidelity from a cold TLB start is considered negligible. \Response{This assumption holds even under hardware virtualization, as the short instruction lengths of 100k instructions require only a handful of pages, allowing execution to reach a steady state within the early phase of the snippet execution. While this particular steady state may not be identically mirroring CPU conditions in silicon, it remains consistent between Panthera and RTL, which is the essential requirement for valid comparison.}

\Response{
It is worth mentioning that performance snippets serve double duty as power collateral. The power validation team relies on real world code for test content, and the PV team has a seemingly unlimited supply! A curated subset of high-power stimuli was selected from the PV test library and provided to the power team. These candidates were then analyzed using the pre-silicon energy infrastructure to identify the highest power tests, with the resulting set of $\sim$15 precious tests used to exercise and validate power delivery through the core. 
}
\section{Unit-Level Verification}

As the complexity of core and SoC modules grows, conducting unit-level PV becomes increasingly valuable. Verifying a module independently of the rest of the system enables focused testing of its specific functionality, allowing for more in-depth examinations of the design and corner cases without interference from other components. In addition, isolating the module simplifies debugging by narrowing the scope, making it easier to identify and resolve issues. PV at different abstraction levels has been shown to enhance efficiency by detecting problems in the smallest unit, prior to integration into a larger system \cite{pv_china}. This layered methodology reduces the overall effort and verification timeline. Here we provide two in-depth studies of unit-level PV at work.

\subsection{Case Study: Branch Prediction Unit}
\begin{figure*}[hbp]
    \centering
    \begin{minipage}[b]{0.24\textwidth}
        \centering
        \includegraphics[width=\textwidth]{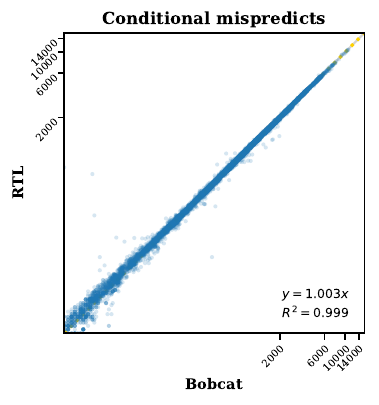}
    \end{minipage}
    \hfill 
    \begin{minipage}[b]{0.24\textwidth}
        \centering
        \includegraphics[width=\textwidth]{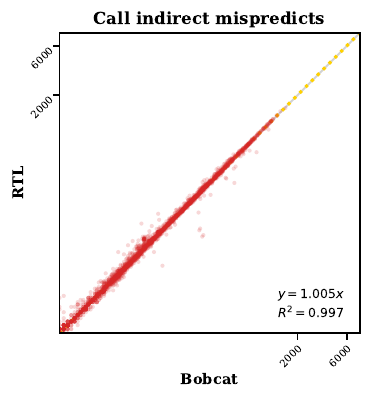}
    \end{minipage}
    \hfill
    \begin{minipage}[b]{0.24\textwidth}
        \centering
        \includegraphics[width=\textwidth]{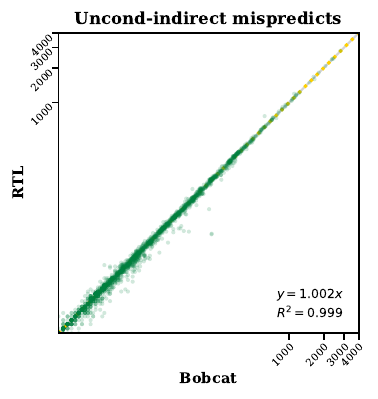}
    \end{minipage}
    \hfill
    \begin{minipage}[b]{0.24\textwidth}
        \centering
        \includegraphics[width=\textwidth]{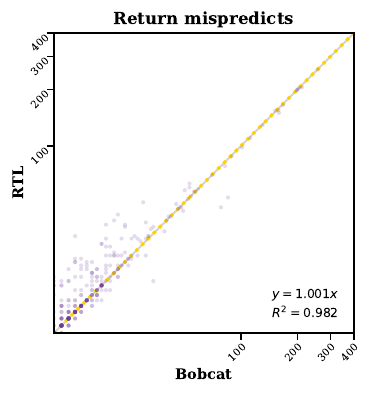}
    \end{minipage}
    \caption{Correlation of the four types of branch misprediction counts. Each plot compares the misprediction count in the RTL versus Bobcat on log scales.}
    \label{fig:bpu_correlation}
\end{figure*}

The BPU is an illustrative example of why this approach is necessary; it is one of the most important pieces of a modern processor because it directly affects the speed at which the front-end delivers the correct work to the back-end. BPU RTL bugs, typically manifesting as a rise in branch misprediction rate, tend to have cascading and compounding effects throughout the pipeline.\footnote{This section assumes familiarity with TAGE branch predictors \cite{tage, ittage}.}

The BPU's position at the very front of the pipeline makes it easier to isolate and test on its own, motivating BPU unit-level test frameworks for both RTL and Panthera. Here, instead of correlating the IPC results of both models, we track metrics tied to BPU execution: namely, \textit{branch mispredictions}.

Panthera's BPU model is wrapped in a smaller framework that we call \textit{Bobcat} \cite{bobcat}. The Bobcat simulator consumes instruction traces, exercises the same BPU modeling code as Panthera, and outputs a statistical breakdown of how the BPU performed, including counts of all fetched branches and mispredictions. 

The BPU RTL unit-level test environment is similarly built to exercise the BPU RTL code. This environment serves as the primary vehicle for DV of the front-end of the CPU, and it doubles as our BPU RTL PV environment. As input, the test environment takes a stream of branches, as well as knowledge of the golden resolutions (targets) of each branch. The output of each test run is a summary of the same branch statistics that Bobcat reports. This unit-level framework simulates instruction fetch magnitudes faster than the full-core RTL design with 100x more coverage.

Philosophically, verification engineers are always looking for useful content to test the designs in meaningful ways. Writing microbenchmarks to stress the BPU can get cumbersome, with an exponential set of possibilities for each predictor and branch history length. Instead, we leverage the curated application snippets from \S\ref{app_snippets} as a large-scale, real-world stimulus. By using all the waves of snippets, each correlation run exercises 2.2 billion instructions and 500 million branches.

Our BPU snippets contain four kinds of branches that can mispredict, each of which can be correlated separately:
\begin{itemize}[leftmargin=*]
\item\textbf{\textcolor{cond}{Conditional}:} Branches that change control flow based on a condition being true or false, such as if-else statements or loops.
\item\textbf{\textcolor{call_ind}{Call Indirect}:} Branches used for virtual function calls or switch statements, where the target address is determined dynamically from a register or memory value. 
\item \textbf{\textcolor{uncond_ind}{Unconditional Indirect}:} Branches that always jump to a target address determined at runtime, stored in a register or memory.
\item \textbf{\textcolor{return}{Return}:} A subset of indirect branches that transfer control back to the instruction that follows a prior CALL.
\end{itemize}
    Misprediction correlation plots for these four types are shown in \autoref{fig:bpu_correlation}. This data is aggregated over 11,000 snippets, each being run for 200,000 instructions. The fundamental goal is to ensure that the same behaviors are seen between RTL and Bobcat. Ideally, all points would be on the $y=x$ line, indicating agreement on the number of mispredictions. Deviations that are far from this line are worthy of further investigation. These could be root-caused to environment issues, or they could be real performance bugs.

A large number of early outliers were a result of key behavioral differences between how the two frameworks process instruction streams. First, the default randomness present in the functional DV environment drastically differs from the deterministic Bobcat runs. To address this, we use environment knobs to intentionally switch RTL into a fully deterministic mode. This disables items such as variable cache miss latencies, random snoop invalidations, and random flushes, and provides consistent execution for correlation. Second, Bobcat models the full prediction, execution, and update of a branch in a single step, all before beginning to predict the next branch. BPU RTL, however, is pipelined; predictions are made speculatively in the shadow of older in-flight branches, whether the older branches were predicted correctly or not. This may result in discrepancies in a branch's predicted target. To reconcile this difference, we implemented a ``nonspeculative fetch'' mode in BPU RTL to mimic Bobcat, in that a branch will only be predicted once the branch fetched before it has fully retired.

Ultimately, the results of our approach were substantial and helped validate our BPU uArch choices. Performance issues were uncovered in both the RTL and in the Bobcat performance model, proving the value of the cross-checking approach. At the end of the project, our BPU RTL matched the architecture model at 99.9\% accuracy, giving the team very high confidence in the prediction quality of the BPU. Just as important, the number of BPU performance issues that were identified during full core-level testing dropped to almost zero, meaning that problems were caught much earlier in the development process. This work saved the team significant effort and demonstrated that a dedicated approach for the BPU is an essential part of core PV. Here are some examples of BPU design bugs that were found and fixed:
\begin{itemize}[leftmargin=*]
\item \textbf{Array sizing}: Basic failures to update the physical sizing of prediction structures based on the latest design specification.
\item \textbf{Buffer allocation}: Mismatching allocation logic for TAGE global arrays based on strength and usefulness.
\item \textbf{Branch aliasing}: Conditional branches were aliased to the same TAGE entry with different directions, ultimately solved by allocating into an array with deeper history.
\item \textbf{BTB capacity}: BTB (Branch Target Buffer) ran out of indirect entry storage in a `way'. Two branches were causing the thrashing: the first was a monomorphic indirect that could get predicted solely out of the BTBs. The second was a polymorphic indirect, but the design incorrectly filtered ITTAGE hits with BTB hits and thus never used the correct ITTAGE prediction.
\item \textbf{Modeling discrepancies}: Return-stack-buffer's full condition wasn't modeled properly in Bobcat. Modeling an infinitely-sized RSB covers 99\% of the behaviors accurately; however, a handful of Java tests did exercise the real depth of the RSB which motivated more accurate modeling. This then enabled discovering more issues in the design.
\end{itemize}

\subsection{Case Study: L2 Prefetcher Subsystem}

Modern high performance cores implement prefetchers for memory accesses, as it is one of the best ways to gain performance. The Ampere core is no exception, and employs a L2 prefetcher subsystem designed to mitigate memory latency. This system integrates four distinct algorithms operating in concert: an implementation of the Best-Offset Prefetcher (BOP) \cite{best_offset_pref}, which dynamically scores a set of candidate address offsets in a ``horse race'' to determine the most effective one; a second-best offset prefetcher, which uses the same scoring and chooses the next-highest-scoring offset; a simple next-line prefetcher; and an adjacent-sector spatial prefetcher for exploiting localized data access. This section details the multi-stage methodology developed to validate the L2P, which required a bespoke debug approach to overcome challenges inherent in timing-sensitive hardware.\footnote{This section assumes familiarity with the BOP prefetcher algorithms \cite{best_offset_pref}.}

\begin{figure}[ht]
    \centering
    \includegraphics[width=1.0\columnwidth]{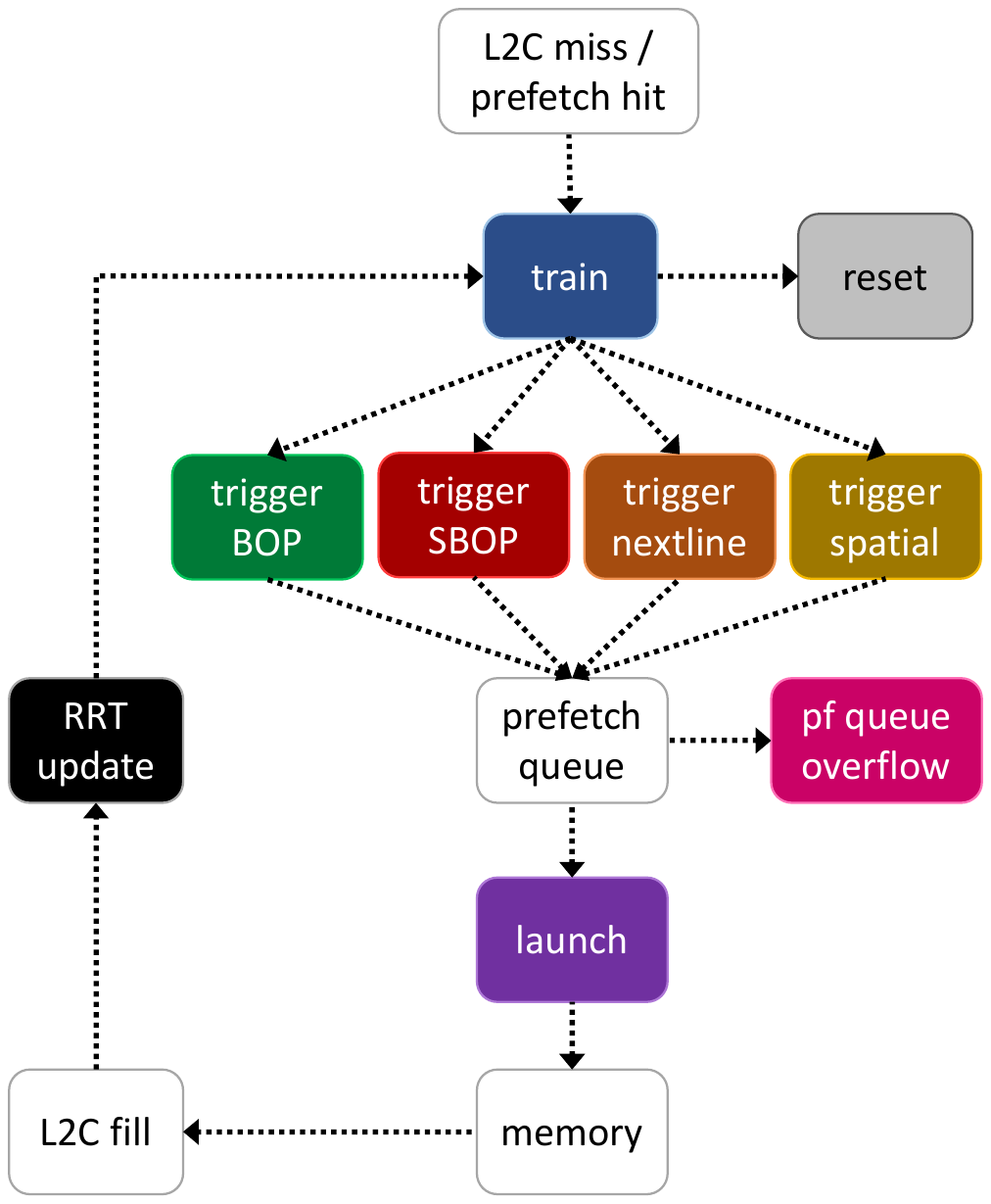}
    \caption{L2 Prefetcher block diagram of behavioral pipeline}
    \label{fig:l2p_block}
\end{figure}

\begin{figure*}[hb]
    \centering
    \begin{minipage}[b]{0.24\textwidth}
        \centering
        \includegraphics[width=\linewidth]{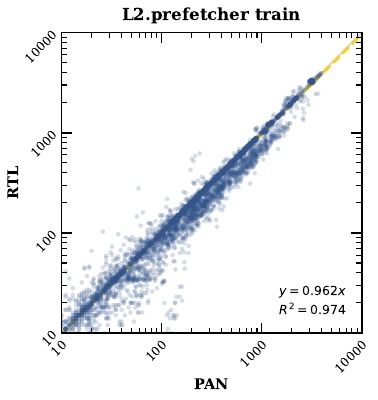}
        \label{fig:l2p_train}
    \end{minipage}%
    \hfill
    \begin{minipage}[b]{0.24\textwidth}
        \centering
        \includegraphics[width=\linewidth]{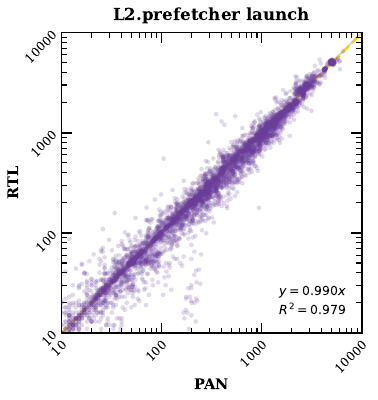}
        \label{fig:l2p_launch}
    \end{minipage}%
    \hfill
    \begin{minipage}[b]{0.24\textwidth}
        \centering
        \includegraphics[width=\linewidth]{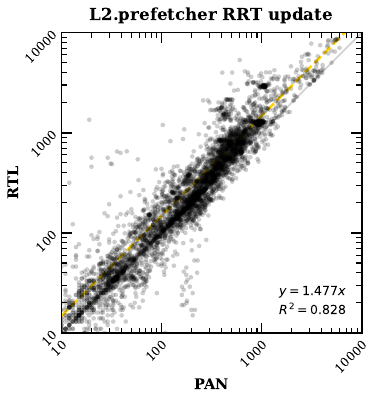}
        \label{fig:l2p_rrr_update}
    \end{minipage}%
    \hfill
    \begin{minipage}[b]{0.24\textwidth}
        \centering
        \includegraphics[width=\linewidth]{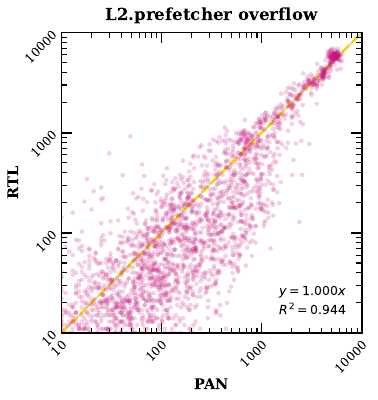}
        \label{fig:l2p_pfq_overflow}
    \end{minipage}%
    \\[-11pt]
    \begin{minipage}[b]{0.24\textwidth}
        \centering
        \includegraphics[width=\linewidth]{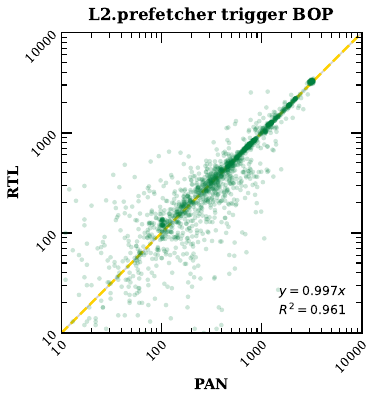}
        \label{fig:trigger_bop}
    \end{minipage}%
    \hfill
    \begin{minipage}[b]{0.24\textwidth}
        \centering
        \includegraphics[width=\linewidth]{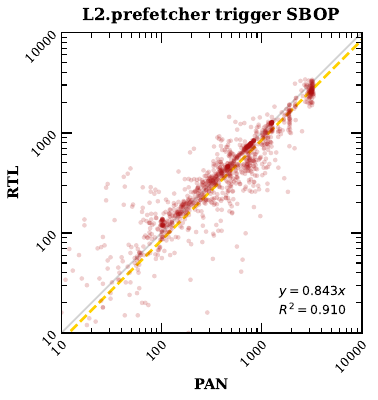}
        \label{fig:trigger_sbop}
    \end{minipage}%
    \hfill
    \begin{minipage}[b]{0.24\textwidth}
        \centering
        \includegraphics[width=\linewidth]{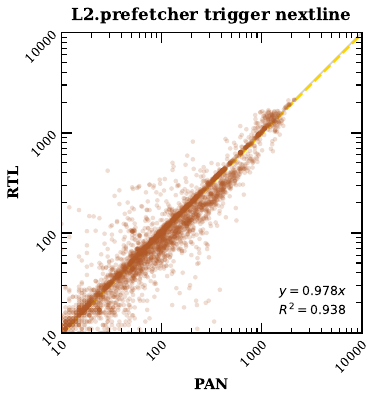}
        \label{fig:trigger_nl}
    \end{minipage}%
    \hfill
    \begin{minipage}[b]{0.24\textwidth}
        \centering
        \includegraphics[width=\linewidth]{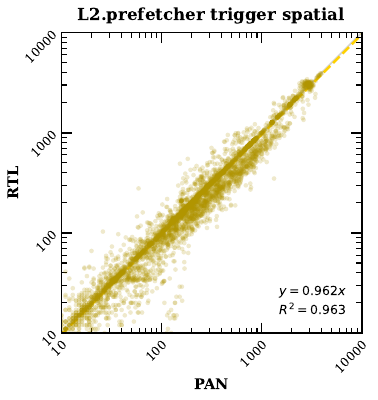}
        \label{fig:trigger_sp}
    \end{minipage}
    \caption{Correlation of key pipeline events for the L2 prefetcher subsystem, and the trigger events for the four L2 prefetchers. Each plot compares the event count in the RTL versus Panthera on log scales. The goal is to have perfect correlation with $y=x$ and $R^2=1$. The most important are the `train' and `launch' events, which depict the input to the pipeline and the output at the end. The poor correlation of `RRT update' motivated the development of the replay mechanism. The colors here match with the diagram in \autoref{fig:l2p_block}.}
    \label{fig:l2_correlation}
\end{figure*}

\subsubsection{Verification Challenge and Initial Strategy}
To enable a disciplined debug process, the prefetcher's behavior was decomposed into a logical pipeline of five key events; the flow can be seen in \autoref{fig:l2p_block}. Events were instrumented in both Panthera and RTL to serve as observation points. These events are:
\begin{itemize}[leftmargin=*]
\item \textbf{\textcolor{train}{Train}:} The learning trigger for the prefetcher, occurring on an L2 miss, or on a demand hit to a previously prefetched line (a ``prefetch hit''). This event serves as a primary input to the learning algorithms, and also reinforces successful predictions.
\item \textbf{\textcolor{trigger_bop}{Trigger}:} The decision by one of the prefetcher algorithms to place a new prefetch request on the output queue, based on its internal state and recent training history. This represents the logical output of the prefetcher's decision-making process.
\item \textbf{\textcolor{launch}{Launch}:} The successful dispatch of a triggered prefetch request from the L2C arbiter to the next level of the memory hierarchy. This event confirms that a prefetch has physically entered the memory system.
\item \textbf{\textcolor{overflow}{PFQ Overflow}:} The prefetch output queue is a FIFO with a relatively small depth. If prefetches cannot be launched, eventually the queue gets backed up, and new submissions into the queue push out the oldest residents. This ensures timeliness of prefetches.
\item \textbf{RRT Update:} An update to the prefetcher's internal Recent Request Table (RRT) upon the arrival of any data fill into the L2 cache. The RRT is an essential state-keeping structure used to track memory access history.
\end{itemize}

The L2 BOP takes a longer time to warm up than other parts of the microarchitecture. So, a challenge emerged as we had to reconcile the L2P's long start up time with the short run-length of RTL simulations which are computationally expensive and limited to 100,000 instructions. Luckily, the BOP is configurable and we have options to adjust the algorithms. To make verification feasible, we created a ``quick-train'' configuration, shown in  \autoref{tab:bop_quick}, which significantly lowered the algorithmic thresholds for training cycles and state resets. Using an analogy of a teenager driving a car for the first time and overcompensating turning the steering wheel, we can also make BOP adjust quickly to actuate its algorithms. When enabling quick-train in both Panthera and RTL, the prefetcher activates and adapts rapidly, ensuring its transitionary behavior is exercised within the limited simulation window. Additionally, we selected only snippets that were known to significantly exercise the L2 prefetcher; core bound tests with no memory traffic are not relevant for this device under test.

\begin{table}[t]
    \centering
     \caption{Parameters modified to enable quick training for PV.}
    \setlength{\extrarowheight}{.3ex}
    \begin{tabular}{lrr}
    \hline
       \textbf{BOP configuration}  & \textbf{default\,perf} & \textbf{quick\,train} \\
       \hline
       max rounds reset & 127 & 15 \\
       max score reset  & 31 & 7 \\
       bad score limit 1 (BOP)  & 2 & 2 \\
       bad score limit 2 (SBOP) & 7 & 4 \\
       \hline
       \\
    \end{tabular}
    \label{tab:bop_quick}
\end{table}

Our initial correlation strategy was to ``peel the onion''—a methodical, stage-by-stage comparison of event counts (\autoref{fig:l2_correlation}). We began by focusing exclusively on debugging mismatches in \verb|train| events across our entire test suite. Once the train counts were well-correlated we moved our focus to the next stage, \verb|trigger|, and continued this process down the pipeline, incrementally identifying and fixing bugs.

\subsubsection{The Replay Mechanism: Decoupling Logic from Timing}
While this staged approach was effective for discovering initial, high-level bugs, its utility diminished as we progressed deeper into the pipeline. We found that even minuscule timing differences in memory return paths between the abstract Panthera model and the cycle-exact RTL would cause chaotic, divergent behavior in downstream events. For instance, there could be a slight reordering events in the update of the RRT (\verb|rrt_update|). This is something that is functionally acceptable in a real system, yet can lead to a different best-offset being chosen by the BOP algorithm. This single change would then cascade into a completely different sequence of trigger and launch events, making it nearly impossible to determine if a discrepancy was due to a legitimate logic bug or simply a minor, \textit{acceptable} timing variation.

To overcome this challenge, we developed a replay mechanism within Panthera. We refactored the BOP model code to consume a pre-recorded event-stream from an RTL run and use it as a direct stimulus. In this replay mode, instead of reacting to its own simulated memory system, Panthera's prefetcher algorithm was driven by the exact sequence and timing of \verb|train| and \verb|rrt_update| events generated by RTL. It was important to get the order correct for all memory events, including speculative path transactions and MMU related requests. This technique was made possible by two key design choices: the clean separation of the prefetcher's core algorithmic logic from timing-dependent code in Panthera, and the use of an identical event-stream format between Panthera and RTL.

The replay mechanism proved to be a breakthrough for this verification effort. It provided a true apples-to-apples comparison of the prefetcher logic by completely isolating the algorithm's behavior from the ``noise'' of the surrounding memory system. This technique facilitated identifying the root-cause of a host of subtle bugs in the RTL implementation, including:
\begin{itemize}[leftmargin=*]
\item \textbf{State Corruption}: An incorrect address hashing function was used for RRT indexing, leading to corruption of history state and thus flawed training.
\item \textbf{Algorithmic Errors}: An incorrect two's complement calculation resulted in improper offset subtractions during training, leading to wrong prefetch decisions.
\item \textbf{Microarchitectural Inefficiencies}: The RTL failed to coalesce multiple pending prefetch requests to the same output address, wasting L2 pipeline bandwidth.
\item \textbf{Resource Hazards}: Prefetches were being dropped during simultaneous read/write accesses to the internal prefetch buffer due to a logic hazard.
\item \textbf{Buffer Management Flaws}: The prefetch buffer allocation logic was inefficient, overwriting valid entries before all invalid entries were utilized.
\end{itemize}

These bugs were either fixed in the design where they were found, or the behavior was studied in Panthera to determine a better solution for the subsequent design. Interestingly, we found a single workload snippet which had a very predictable streaming memory access pattern, combined with enough computation in between each request, to make prefetching supremely useful. This one test proved particularly effective at exercising all of the prefetcher's core functionalities, and was instrumental in uncovering many of the issues cited above. 

\autoref{fig:l2_correlation} shows the correlation of pipeline events: the inputs (\verb|train|, \verb|rrt_update|), and the outputs (\verb|launch|, \verb|overflow|) as well as triggers for the four subcomponent prefetchers. This data was collected from 7,000 snippets near the end of PV for the first core product. The PV efforts for the subsequent core projects tightened the correlation even further.

Overall, the trendlines show correlation close to $y=x$. Panthera tends to have more trigger events because there is less interference from wrong path transactions in the RRT. We can see that RTL has consistently higher \verb|rrt_update|'s overall, which leads to more bad-score shutdowns (not shown here). Ultimately, we seek to correlate on \verb|launch| events absolutely, as well as IPC uplift through divergence debugging (\S\ref{sec:divergence_debugging}). Since the correlation and IPC are within the bounds of acceptance, the team declared logical correctness and performance fidelity on the L2 Prefetcher and moved onwards.

\section{Full-Core Correlation}

In concert with unit-level testing, we correlate the ``full-chip'' core RTL with the Panthera core model running application snippets. When all the individual units are integrated and working in concert, new interactions inevitably expose issues not seen before at the unit-level testing. And the knowledge of clean unit-level tests provides the confidence needed to peel the onion further.   

\subsection{Methodology}

RTL is cycle-exact by definition, and the implementation typically runs a few months behind the simulator, as the simulator usually serves as the first stage of a proof-of-concept. Panthera is cycle-accurate, or perhaps cycle-approximate, as there are varying degrees of abstraction modeled in the different units due to the organic nature of increasing fidelity only when needed to improve accuracy. The development effort is guided by keeping key latencies and structure sizes aligned with RTL, while implementing the control algorithms in a behavioral style. For instance, we employ a variety of memory models, some more abstract than others. In PV, we use a fixed memory latency for L2 cache misses that is the same between Panthera and RTL. Near the end of the project, we change this fixed latency to different values to make sure the RTL design still correlates well. Similarly, while we typically keep structure sizes fixed, we support parameterization of some structure sizes, in order to better stress the design. For example, correlating on a small BPU configuration helps to thoroughly exercise aliasing and thrashing scenarios.

The RTL has its own dedicated (functional) test suite including general directed tests, feature-specific directed tests, and constrained-random tests to achieve broad coverage even in corner cases. To maximize flexibility, all microarchitectural features are designed to be as configurable as possible, in both RTL and Panthera. When PV begins, all discrete features are disabled in the configurations and are only enabled once the corresponding RTL implementation has been confirmed to be ready to exercise. This ensures the comparisons between the two are equal, since each is being continually developed.

\subsection{Correlation Metrics and Acceptance Criteria}\label{sec:correlation_metrics}

To ensure a quantifiable and data-driven PV process, we chose a set of metrics to track the correlation between the RTL design and the Panthera model. These metrics serve as the basis for our acceptance criteria, providing targets for the PV team and a definitive measure of when the design's performance is considered validated.  We plot these metrics on a web dashboard as seen in \autoref{fig:dashboard}.

The performance delta for any given test is the ratio of IPC between the two models: $\text{IPC}_{\text{RTL}} / \text{IPC}_{\text{Panthera}}$. From this, we derive our primary aggregate metrics:

\begin{figure}[tbp]
    \centering
    \includegraphics[width=0.92\columnwidth]{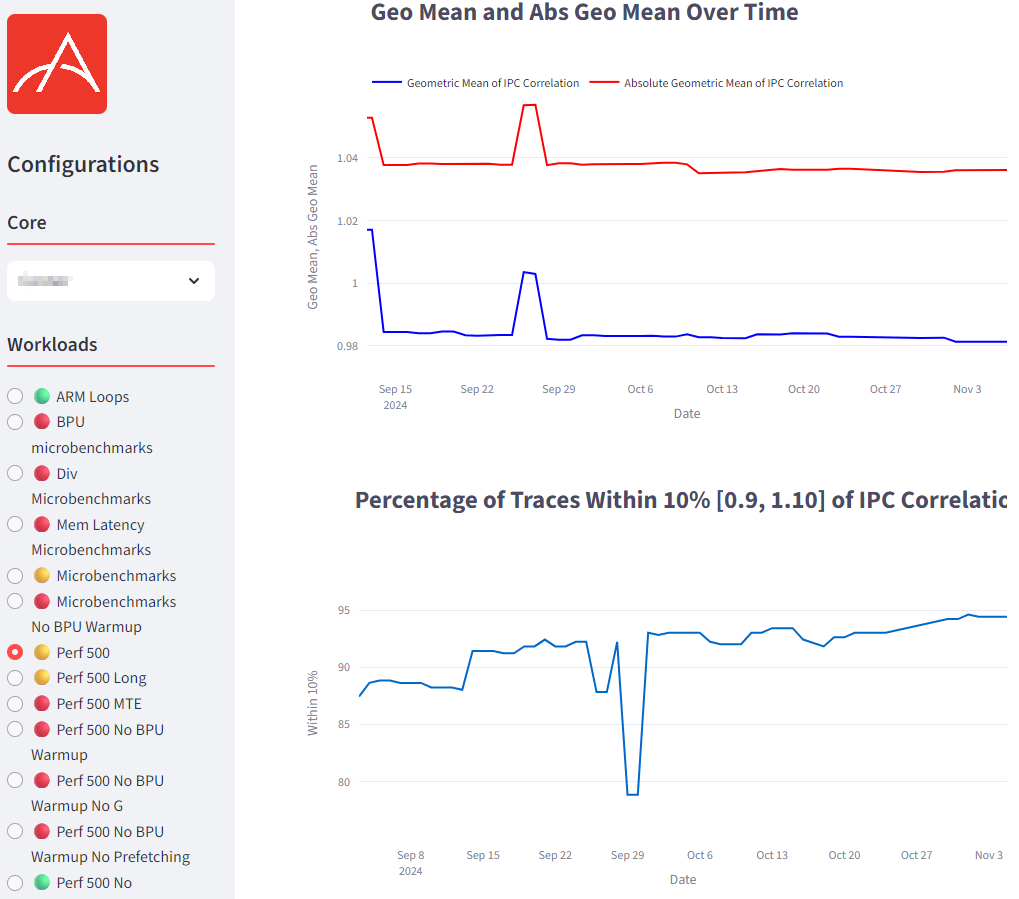}
    \caption{Snapshot of the Daily Dashboard. The correlation metrics are plotted over time, and the data is keyed by different waves of workloads. The GUI makes it easy for the team to track health across products and across tests.}
    \label{fig:dashboard}
\end{figure}

\begin{figure}[tbp]
    \centering
    \includegraphics[width=0.92\columnwidth]{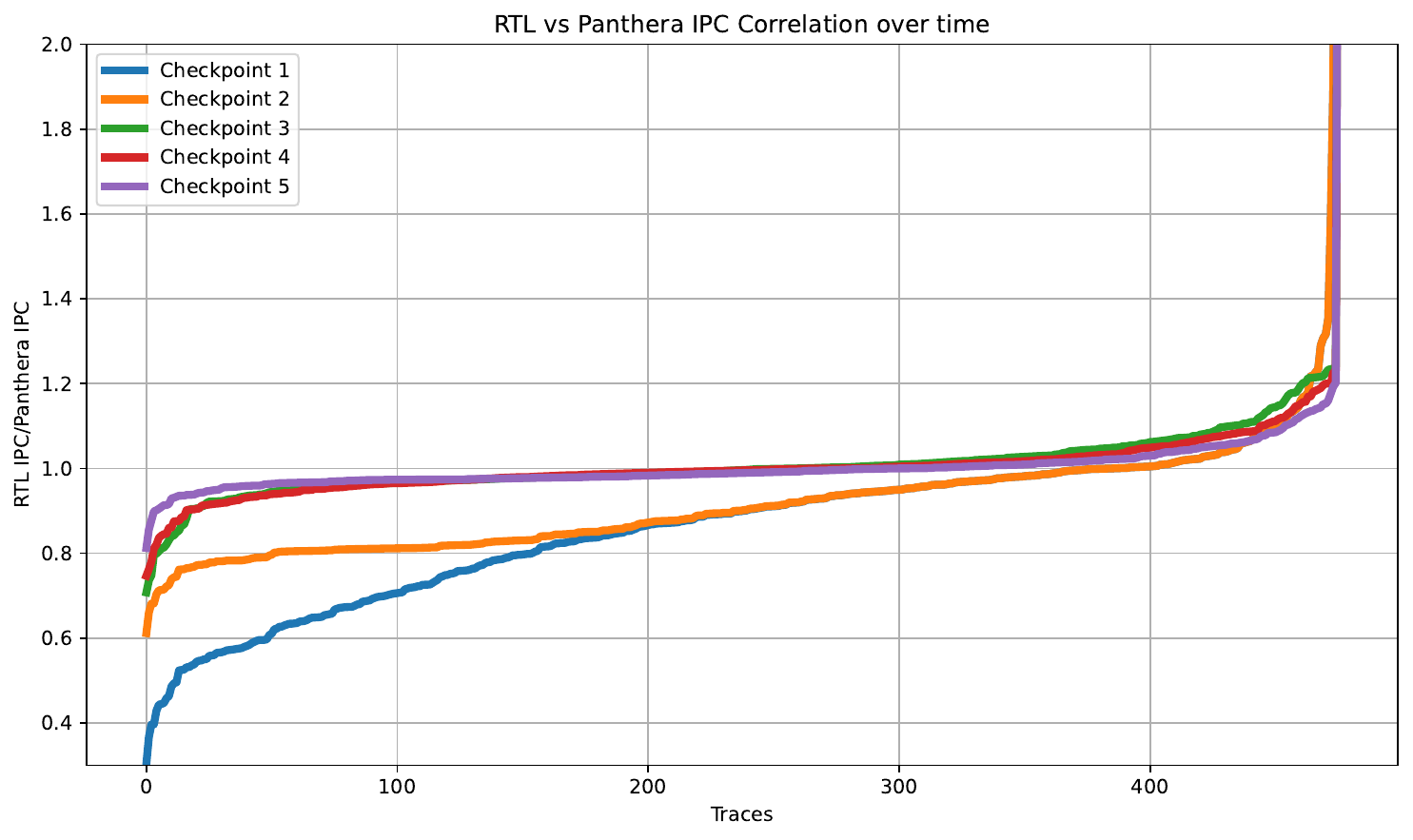}
    \caption{Workloads Snippets IPC S-Curves through time. In the due course of performance correlation, the s-curve starts out far from perfection. The first checkpoint shows a very steep slope on the left, indicating many of the tests are slower on RTL than on Panthera. Debugging these outliers and fixing bugs helps to ``prop up'' the left side of the s-curve, which may also lift the right side (meaning a new Panthera issue is exposed). The ultimate goal is to have 90\% of the tests within 10\% (between 0.9 and 1.1).}
    \label{fig:s_curve_column}
\end{figure}

\begin{itemize}[leftmargin=*]
    \item \textbf{Geometric Mean of Ratios:} The geometric mean of the IPC ratios across our entire suite of thousands of tests is our top-level indicator of overall correlation. A value of 1.0 indicates perfect correlation on average. To visualize the distribution and easily identify outliers, the individual IPC ratios are sorted and plotted, creating a characteristic S-curve (\autoref{fig:s_curve_column}) that reveals the range and severity of discrepancies. 

    \item \textbf{Absolute Geometric Mean:} A standard geometric mean can be misleading, as a large positive deviation (e.g., 2.0x) can be mathematically masked by a large negative deviation (e.g., 0.5x), resulting in a mean close to 1.0 despite significant errors. To prevent this, we also employ an ``absolute'' geometric mean. For this calculation, any ratio less than 1.0 is inverted (i.e., $x \rightarrow 1/x$) before the mean is computed. This metric is insensitive to the direction of the error and provides a true measure of the average discrepancy magnitude.

    \item \textbf{Percentage Box:} While the aggregate means provide a high-level summary, a more intuitive measure is the percentage of tests that fall within a specific error bound. This metric forms our primary exit criterion. For the initial core design, our goal was to achieve \textbf{80\% of tests within $\pm$10\%} correlation bound. For each subsequent product generation, this target was tightened, with the current goal being \textbf{90\% of tests within $\pm$10\%}. 
\end{itemize}

It is important to note that these strict criteria are sometimes adjusted for specific classes of workloads. For instance, tests with heavy operating system (EL1) activity are difficult to correlate with high precision due to the complex modeling required for memory barriers and fences. For these waves of tests, a wider acceptance bound is pragmatically applied.

Finally, all performance analysis is predicated on a functionally stable design. Therefore, a baseline functional health metric is also a key component of our criteria. The RTL execution run must execute and commit the entire flow of instructions in the test for that single test to be considered successful; partial runs are not compared. We require a minimum of 95\% of all performance tests to execute without functional errors in RTL at the beginning of the correlation phase, with a strict exit criterion of \textbf{100\% functional health} by the end of the project. This ensures that any observed performance deviations are genuine microarchitectural discrepancies, not artifacts of infrastructure issues or functional bugs.

\subsection{Modeling Wrong-Path Execution}
Accurately modeling the performance impact of wrong-path execution is a well-known challenge in processor simulation~\cite{correct_wrong_path}. Although these transient instructions are flushed before retirement, they create contention for microarchitectural resources such as caches, prefetchers, and branch predictors~\cite{ooo_wrongpath}. This effect has a dual nature: wrong-path execution can be beneficial if it speculatively fetches data or instructions that are later used by the correct path (constructive prefetching), but it can also be detrimental if it evicts useful data that must then be re-fetched from memory (destructive eviction). The magnitude of this potential performance delta increases in modern, deeply out-of-order processors with large instruction windows and high memory latencies.

As a trace-based simulator, Panthera does not execute instructions and therefore cannot organically generate wrong-path effects, as the trace only contains the committed, correct-path instruction stream. To overcome this limitation, our trace collection methodology saves off enough memory information to help model speculative execution paths. By leveraging methods analogous to resurrected code~\cite{resurrected_code} and instruction reconstruction~\cite{inst_reconstruct}, we capture the likely targets and outcomes of mispredicted branches. Using this control flow, the last-known right-path register values, and known instruction text, we perform best-effort emulation of wrong-path instructions to generate register values needed to calculate the addresses of wrong-path memory accesses. This allows the Panthera model to emulate the memory access patterns of speculative path instructions, and pollute the cache hierarchy in a manner that approximates the behavior of RTL. While no trace-based approach can achieve perfect fidelity, the integration of wrong-path modeling capabilities showed a measurable improvement in model correlation. Across our test suite, we observed a \textbf{1.3\% improvement} in the absolute geometric mean, a \textbf{4.4\% increase} in the number of traces correlating within our $\pm$10\% target bound (see \S\ref{sec:correlation_metrics}), and over \textbf{50\% reduction in error for extreme outliers}.

\Response{
The wrong-path modeling was specifically motivated by PV's 100k instruction length, and is not needed in pathfinding or projection's 10M+ instruction length. We have shown good correlation between the Panthera model (without wrong-path) against AmpereOne\textsuperscript{\textregistered} silicon with 192 cores active \cite{ampere_mav}. Although we could enable wrong-path for full-length studies, it is not needed. The average impact of wrong-path in these cases is negligible, so there isn't much fidelity loss from wrong-path loads with a trace-based model. Additionally, the traffic shaping algorithms with a full-system load deprioritize the prefetchers in the memory subsystem. We should acknowledge that an SoC with few cores would be more susceptible to this issue than one with many cores.
}

\begin{figure}[b]
    \centering
    \includegraphics[width=0.7\columnwidth]{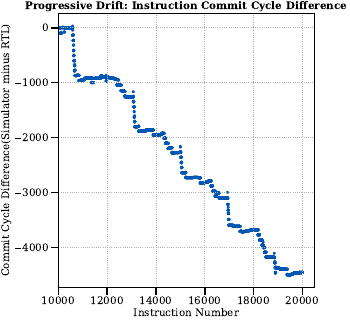}
    \caption{A plot of cumulative cycle drift vs instruction count for a single outlier test. The first 10,000 instructions of execution are used to warmup the pipeline on both Panthera and RTL, and the subsequent 10,000 are the measurement region. Flat regions indicate the behaviors match exactly between the design and the model. The majority of drift is punctuated in five small regions (e.g at 10,500 and 13,000). Therefore, we can focus our debug in those regions to isolate the cause of miscorrelation.}
    \label{fig:progressive_drift}
\end{figure}

\subsection{Event-Stream Based Analysis}

The foundation of our quantitative analysis is a unified event-stream framework, which provides the raw data for both high-level metric calculation and deep-dive debugging. During a simulation, both Panthera and RTL will emit a time-ordered log of key microarchitectural occurrences into an ``event-stream'' file, in a well-defined \verb|protobuf| format \cite{protobuf}. Each event in this stream consists of a timestamp, an event type (e.g., an instruction decode), and a set of key-value pairs containing event-specific data.

From these event-streams, all performance metrics are calculated. For example, the L2 cache misses per thousand instructions (MPKI) is derived by simply dividing the total number of ``L2 cache miss'' events by the total number of ``Instruction Commit'' events. Typically, only events occurring within the designated measurement phase of a simulation (after warm-up and before cool-down) are included in this calculation, to ensure equality of work between models operating in steady-state regions.

\begin{figure}[tbhp]
    \centering
    \includegraphics[width=\columnwidth]{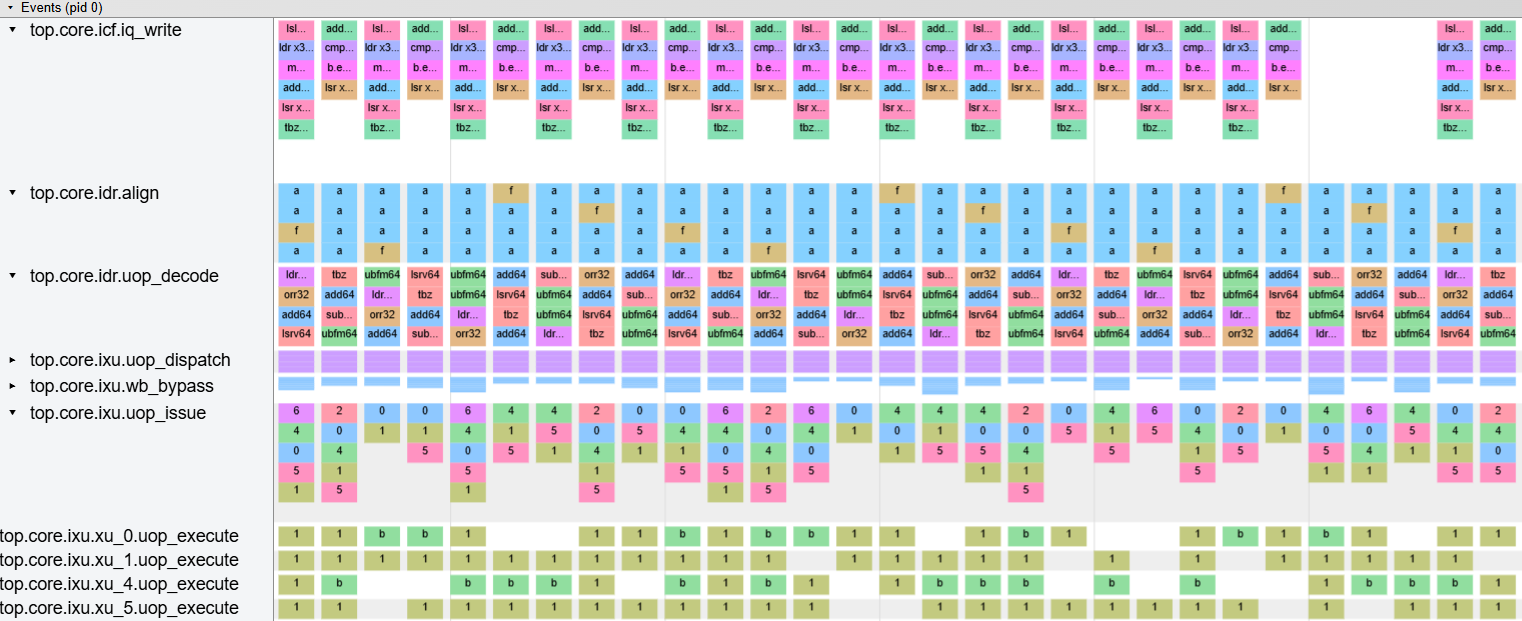}
    \caption{A snapshot of an event-stream visualized using the \textit{Chromium Catapult} interactive trace-viewer \cite{catapult_viewer}, with a resource-oriented view with cycles on the x-axis and various microarchitectural events on the y-axis. The sample above shows the core pipeline running in a high IPC region of code, with the front-end delivering four uops per cycle to the scheduler and execution units.}
    \label{fig:catapult_pipeline}
\end{figure}

\begin{figure}[thbp]
    \centering
    \includegraphics[width=\columnwidth]{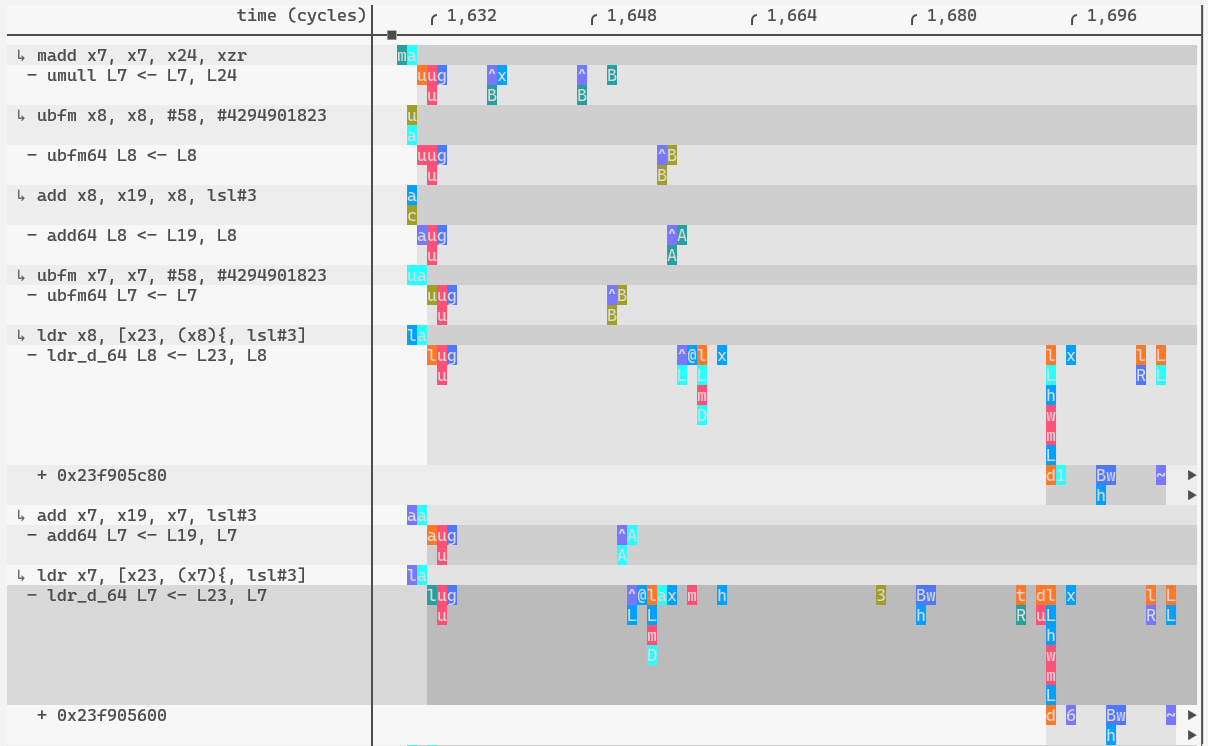}
    \caption{A snapshot of an event-stream visualized using \textit{Catscan}, an interactive trace-viewer with a transaction-oriented view showing instructions, uops, and cache misses. We see two load uops that each miss the L1 cache \textbar{}\misshighlightnarrow{m}\textbar{}, spawn transactions by allocating fill buffers for demand requests \textbar{}\allochighlightnarrow{d}\textbar{}, that hit in the L2 six cycles later \textbar{}\hithighlightnarrow{h}\textbar{} and eventually return data to the L1 and deallocate  \textbar{}\deallochighlightnarrow{\raisebox{-0.80ex}{\textbf{\textasciitilde}}}\textbar{}.}
    \label{fig:catscan}
\end{figure}

Beyond aggregate metrics, these event-streams are the primary tool for root-cause analysis of performance discrepancies. The debugging workflow begins by analyzing high-level performance plots to identify the precise timestamp where a deviation originates. For instance, in the example shown in \autoref{fig:progressive_drift}, the first performance cliff is clearly visible around instruction 10,500. An engineer then inspects the raw event-streams from both Panthera and RTL around this point, using tools such as those seen in \autoref{fig:catapult_pipeline} and \autoref{fig:catscan}, which allow the architect drill down into event causalities to debug pipeline hazards and behavioral microarchitecture bugs.

Our implementation strategy for this framework follows a model-first approach. New events are typically first created in Panthera, where the higher level of abstraction allows for rapid development and validation. A targeted subset of these events is then ported to the RTL simulator on an as-needed basis for specific debugging tasks. This selective approach is a pragmatic necessity; the higher fidelity of the RTL makes event instrumentation complex and resource-intensive. The integrity of this entire methodology hinges on this porting process, as it is \textit{critical} that an event of a given name has the exact same semantic meaning in both the model and the RTL.

Our methodology is centered on a high-frequency regression model for continuous correlation and rapid bug detection. Central to this process is a daily regression that executes a core set of tests against the latest RTL build. The results are automatically aggregated and plotted on a tracking dashboard, an example of which is shown in \autoref{fig:dashboard}. This provides an immediate, at-a-glance view of the design's performance health over time.

The primary benefit of this daily cycle is the dramatic reduction in time-to-isolation for new bugs. Any sudden divot in a metric's trend line immediately signals a performance regression, and its cause can be quickly isolated to a code change submitted within the preceding 24-hour window. These daily checks are complemented by more comprehensive, full-suite regressions run on a weekly basis, which provide deeper coverage across more waves of workloads. This tiered approach provides both rapid, actionable feedback for developers and deep, systematic validation of the design.

\subsection{Divergence Debugging}\label{sec:divergence_debugging}
\begin{figure}[ht]
    \centering
    \includegraphics[width=0.7\columnwidth]{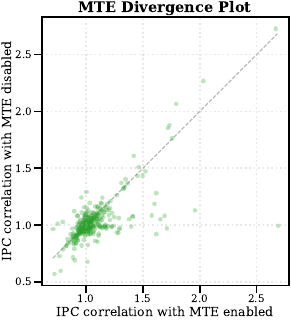}
    \caption{Divergence Plot, on Ampere's implementation of ARM's Memory Tagging Extension running with snippets on RTL.}
    \label{fig:mte_diverge}
\end{figure}

Another technique we use to expose issues is divergence debugging \cite{relative_debugging, divergence_debugging}. One form of this is to measure IPC correlation by toggling a feature on and off. Traces that show strong correlation when the feature is disabled but diverge when it is enabled become prime candidates for further investigation. \autoref{fig:mte_diverge} showcases this technique on Ampere's implementation of ARM's Memory Tagging Extension (MTE) \cite{ampere_mte}. When correlation is perfect with both MTE enabled and disabled, there is no problem; these are the points near (1.0, 1.0). The data points on the $y=x$ line have common mode miscorrelation issues which are not related to MTE. The interesting tests to investigate then, are the ones near the line $y=1$, which have great correlation when MTE is disabled, but poor correlation when MTE is enabled. The point at (2.7, 1.0) is a perfect candidate--we know that the MTE feature is the cause of the issue, and can focus our attention there.

Another example of divergence debugging is to test prefetchers. Since prefetcher behaviors can introduce some natural chaos into correlation work, we can use divergence testing to isolate the issues. The various different prefetchers can be turned on and off, in both RTL and Panthera, to narrow the culprit of miscorrelation.

\subsection{PMU Events Verification}
The accuracy of Performance Monitoring Unit (PMU) events is extremely important, since these are the primary method of post-silicon performance measurement and even functional debug. Traditional methods of functional verification do not carry-over well for PMU verification; each event needs to be stressed, sometimes with custom per-event stimulus. Sometimes randomized testing is able to actuate counters, but many times it does not. This challenge compounds as modern CPUs have several hundred PMU events that require  directed testing. 

To manage this effort, we integrated a PMU events verification methodology into the PV infrastructure. For each PMU event implemented in RTL, we add an event in Panthera with behavior matching the architectural spec. As we collect data for PV verification, we also collect PMU event counts from both RTL and Panthera. For each event, we generate an S-curve, correlation plots, and trace-based rollups. This allows us to quickly identify PMUs that diverge in correlation, identify traces that capture those divergences, and debug and root-cause implementation issues in RTL, piggybacking on the same tools that we use for performance verification.

This methodology has resulted in catching several bugs, and has significantly improved accuracy of PMU events. Since it scales well, it has allowed us to add new events without paying a high verification effort cost per-event. Ampere's fourth generation core has over 500 verified PMU events. 

\subsection{Power-down Override Testing} 
In modern CPUs, power consumption is a critical design constraint, influenced by factors like thermal management and operational costs in data centers. One power saving technique that reduces dynamic power is clock gating. By intelligently disabling the clock signal to inactive or idle functional blocks, clock gating seeks to mask off toggling of sequential elements and combinatorial logic within those blocks. 

A fundamental principle for power-saving clock gating is that it must have zero impact on performance. To ensure this, we leverage PV methodology. Each major block of logic has a clock gating override bit, which can be used to exercise the block with and without clock gating. We run a set of workloads in these two configurations, and check the performance. Ideally, the IPC from these two runs should be identical, showing zero performance delta. 

\begin{figure}[htbp]
    \centering
    \includegraphics[width=0.7\columnwidth]{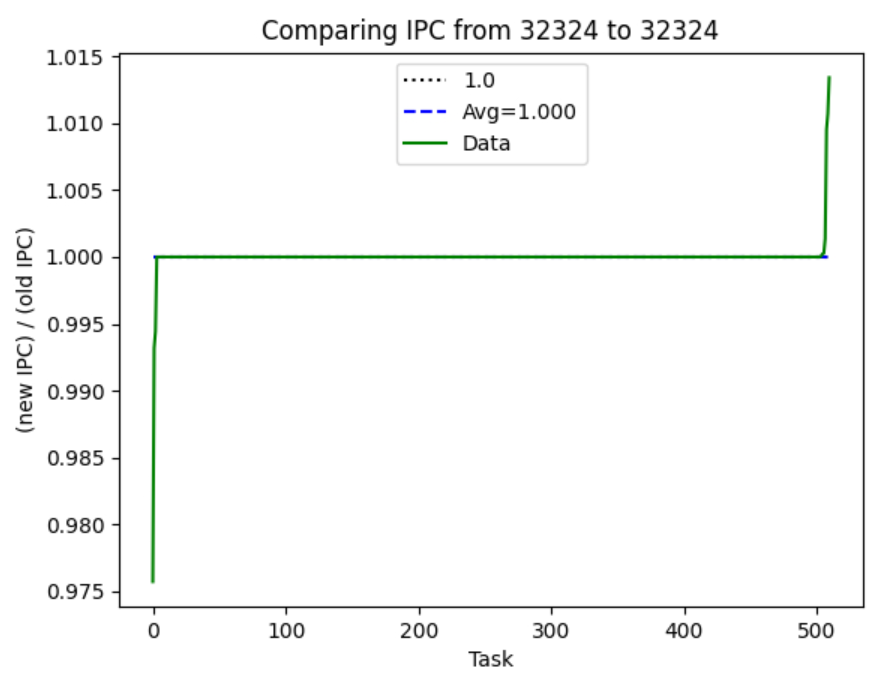}
    \caption{An s-curve across 500 performance snippets comparing clock-gating enabled versus disabled. We see perf deltas on the two ends of the curve, indicating a logic divergence. This case was root-caused to a ping-pong arbiter in the L2 fill selection logic that was being predicated on a power-down select pin, when it should have been running based on fill eligibility instead.}
    \label{fig:pv_pwrdwn}
\end{figure}

Should any discrepancies arise, we take the largest offending outlier and start a binary search on the hundreds of power down override configuration bits. This will determine which cone of logic is responsible for the unexpected performance drift. Then we call in the expert in that area to debug the issue, making the best use of human time. \autoref{fig:pv_pwrdwn} explains how these runs caught a logic bug.

Traditional DV practices such as n-compare, x-injection on clock override signals, and formal methods are used to validate clock gating functional correctness, but each have their own limitations. The PV method shakes out the power-down divergences in the tests where they would be most exposed anyway: the performance snippet collateral. Thus, DV and PV work in concert to provide confidence in the design.

\subsection{Triage of Performance Sightings}

With a continuous stream of performance data being generated, an efficient triage process is needed to manage the debug workload across a team of domain-expert architects. The primary goal of triage is to accurately assign each new performance outlier to the engineer most qualified to investigate it. While this was historically a manual process, the scale and complexity of modern CPU verification motivate an automated approach to ensure good use of human engineering time for root-cause analysis.

Our automated triage pipeline is designed to pinpoint the most likely origin of a performance deviation. The process for each outlier test is as follows:

\para{Metric Filtering} The process begins by pruning the dozens of collected performance metrics to a subset most relevant to the specific test being analyzed. Metrics are programmatically ignored if their values are ``uninteresting'' for a given trace, using a combination of absolute, median-relative, or max-relative filtering methods. This step focuses the analysis on only the most significant signals.

\para{Identifying the Reason Metric} The values of the remaining relevant metrics are then normalized by their statistical variance to ensure a fair comparison between metrics of different scales. A difference or ratio is calculated for each normalized metric against its expected value from the performance model. The metric exhibiting the largest deviation is designated as the ``Metric Guess,'' as it suggests the primary indicator of the bug's underlying cause.

\Response{While such policy is not guaranteed to be accurate, the ``Metric Guess'' was usually \textit{related} to the underlying cause which helps in assigning an owner and providing a starting point for debug. In some cases a better metric was not chosen due to the filtering phase which is the risk of having a generalized filtering heuristic for all metrics. Similarly, sometimes downstream metrics would be selected because of \textit{upstream} reasons, but both are weighted equally and the selected metric had a larger difference. Additionally, in some cases, better metrics which exist in Panthera, do not exist in RTL and thus cannot be chosen.}

\begin{figure}[b]
    \centering
    \includegraphics[width=1\linewidth]{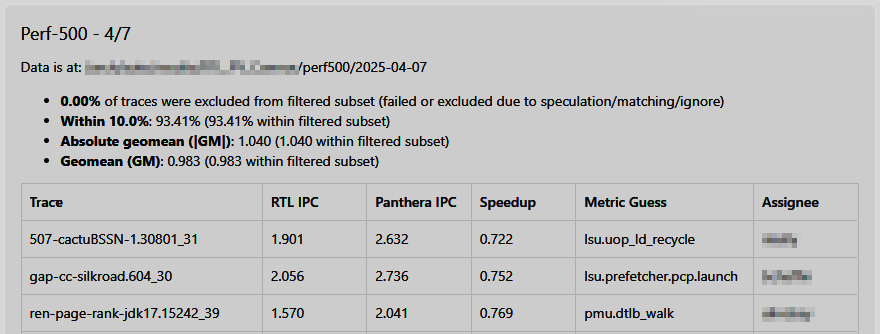}
    \caption{MSTeams Post with debug assignments from the perf-check-500 trace list.}
    \label{fig:triage}
\end{figure}

\para{Automated Assignment} In the final step, the guessed metric is used to assign a bug owner. A predefined mapping connects each potential reason metric to its corresponding microarchitectural unit. For example, a high L2 Cache miss rate without many prefetch requests maps directly to the L2 Cache; while lots of prefetch activity in concert with L2 Cache misses would map onto the L2 Prefetcher unit. The scripts collect all the necessary run artifacts for debug, and then post is made on the RTL PV channel on Teams, containing high level information on the outlier and location of debug collateral, so the engineer assigned to debug can begin work. A sample screenshot is shown in \autoref{fig:triage}. This automated triage system transforms a raw list of performance outliers into a prioritized queue of debug tasks assigned directly to relevant experts, which improves the efficiency and throughput of the entire team.

\Response{
\subsection{Correlation Trends}

Our experience is that snippet correlation is less a function of the high-level application domain and more a function of two key properties: the microarchitectural specificity of the code snippet and the verification maturity of the hardware features it exercises.

Simple code snippets that stress a specific feature (like load-store forwarding) tend to correlate either perfectly or are extreme outliers, making them great for targeted debugging. In contrast, more complex snippets with a myriad of behaviors may show good correlation due to the averaging effects of exercising many different code paths, making them less useful for isolating issues (except perhaps for front-end instruction delivery).

Subsequently, a workload's correlation is a dynamic measure that reflects the verification maturity of the design at a given point in time. For example, a snippet dominated by branch instructions will exhibit poor correlation early in the project when the BPU model is still converging with the RTL. However, once the BPU is stabilized and verified, the same snippet becomes one of our highest-correlating test cases. The correlation is therefore tied to the project phase and the feature under test, not the intrinsic application type.

This understanding informs our test generation strategy: ideal snippets are those that function as microbenchmarks. Extracting these via k-means from real applications is advantageous because it automates the discovery of thousands of relevant, real-world test cases that we may not have thought to write by hand.
}
\Response{
\section{Performance Escapes}

Despite a rigorous PV process, the complexity of a modern CPU means that some bugs may go undetected until post-silicon validation. These ``escapes'' provide lessons for refining and improving future verification plans. Analyzing why a bug was missed is as important as fixing the bug itself. Here we detail all of the performance-related escapes encountered during the development of our custom cores, which led to improvements in our test collateral and methodology.

A significant performance escape in the first-generation silicon was a logic bug in the BTB. The program counter value was not consistently sign-extended, which resulted in mispredictions for any code executing in the upper regions of the address space, such as kernel or system code (EL1) where PCs often begin with \texttt{0xffff}... This issue manifested as an unexpected drop in branch prediction accuracy for system-intensive workloads. The root cause for this escape was a gap in our initial test suite: at that early stage of the project, our tracing infrastructure and workload snippets were focused on user-space (EL0) behavior, and comprehensive EL1 test coverage had not yet been developed.

In the same generation, the Second-Best Offset Prefetcher (SBOP) was found to be inadvertently disabled in hardware under certain conditions. This was a functional bug: when the core switched between secure and non-secure execution states, the logic failed to properly re-enable the SBOP. Although the root cause was functional, the symptom was purely a performance loss. This bug escaped detection because the scenario fell into a seam between two distinct verification efforts. The functional verification plan did not specifically check for performance feature state across security transitions, and the performance verification plan did not include tests that involved switching security modes.

In a later generation core, the initial hardware implementation of MTE suffered from a performance escape where the load-to-store forwarding path was significantly hampered for MTE-related operations, as detailed in prior work (\S4.4.2 of \cite{utaustin_mte}). Similar to the BTB escape, the reason for this was a lack of specific test collateral. The initial PV regression suite did not yet include a sufficient number of tests that exercised the new MTE functionality, leaving this performance-degrading interaction undiscovered.

These cases highlight a common theme; performance escapes are often caused by gaps in test stimulus rather than flaws in the correlation methodology itself. Each of these discoveries led to an expansion of our test suite—adding comprehensive EL1 and system-level traces, incorporating tests with complex mode transitions, and ensuring that new architectural features like MTE have dedicated PV test suites from day one. This continuous feedback loop, turning post-silicon escapes into pre-silicon prevention, is fundamental to the maturation of a robust verification process.
}
\section{Future Directions}
\label{sec:future_work}

The historical trajectory of PV has been one of increasing automation, moving from manually intensive efforts to more systematic, tool-driven methodologies as presented in this paper. Recent publications have systematized the aspect of bug triage for classifying outliers into design units without the use of a golden performance model \cite{bug_triage_md}. Newer techniques in this area include using machine learning techniques to inject performance issues, to then localize performance faults to certain design units \cite{ml_perfval}. The same authors propose focusing on microarchitectural bugs (as opposed to logic bugs) using a system that assumes the existence of a baseline CPU with zero bugs \cite{auto_perf_bug}. Other teams have utilized event counting methodologies, analyzing the output using AI techniques to identify anomalies \cite{ai_perfval, riscv_emu}. While these contributions are valuable, they sometimes rely on simplifying assumptions (such as the existence of a bug-free baseline processor, which is an ideal \textit{rarely} achieved in practice). As the industry stands on the cusp of a new wave of automation driven by generative AI and agentic workflows, it is time to envision a future that is not only more efficient but also preserves the human ingenuity that drives uArch innovation.

Our vision is not one of complete automation, but rather an \textbf{architect-centric agentic workflow}. \Response{Agentic flows have shown promise for design space exploration \cite{agentic_arch, archagent, arch2, archgym, alphazeromoment, berkeley_chia}, and we feel these can be extended to verification as well.} The goal is to leverage AI to handle the laborious, repetitive, and time-consuming aspects of verification---the sifting, sorting, and brute-force testing---thereby liberating architects to focus on the creative, high-impact work of root-cause analysis and next-generation design.  In essence, AI agents should manage the proverbial search for needles in the haystack, so human experts can spend their time understanding \textit{why} a particular needle is interesting and what it implies for future designs. After all, the most fruitful microarchitecture ideas for tuning our Core roadmap have come from humans debugging outliers in PV. To this end, we propose these concrete research directions:

\para{Proactive Anomaly Detection and Triage} Our current daily regression dashboard (\autoref{fig:dashboard}) provides a passive view of performance trends. We envision an AI agent that actively monitors this dashboard. Upon detecting a regression, the agent would autonomously initiate a series of secondary diagnostic runs, such as bisecting the responsible code change or beginning a targeted parameter sweep. By analyzing historical data, the agent could present the human architect with an outlier as well as a high-confidence, pre-triaged bug report that already narrows down the potential cause, transforming the starting point of a debug session from ``what is wrong?'' to ``why is this specific interaction failing?'' In this vein, new research shows an LLM can be trained on memory streams to understand access patterns and cache behaviors \cite{cachemind}, allowing operators quick access to microarchitectural insight (through natural language!) that used to take weeks of analysis.

\para{Exhaustive Microarchitectural Exploration via Parameter Sweeping} While the divergence debugging technique described in Section~\ref{sec:divergence_debugging} is quite effective, it is currently limited by the human capacity to manage parameter configurations and analyze results. An AI agent could systematically sweep through the entire combinatorial space of feature parameters across thousands of tests. This would uncover complex, multi-feature interactions that are currently impractical to find and provide a comprehensive map of the design's behavioral landscape, highlighting areas of unexpected sensitivity or instability. The Ax adaptive experimentation platform \cite{ax_dev} is used in industry 
for managing and automating parameter sweeping explorations using Bayesian Optimization; we have found it a productive tool for post-silicon tuning, and it looks promising for use in PV.

\para{Deep Anomaly Detection with Advanced Metrics} Our future work also aims to move beyond a single correlation metric reason. We plan to integrate more granular metrics into our automated framework, such as those derived from Top-Down Analysis (TDA) methodologies \cite{yasin2014topdown}. An AI agent could be trained to recognize the ``fingerprints'' of specific bug classes within these rich, multi-dimensional data streams. This would allow the system to automatically classify regressions into categories (e.g., ``Front-End Stall due to BTB capacity,'' or ``Memory Bottleneck due to excessive prefetching'') and provide a much more detailed initial diagnosis than top level metrics alone can offer. 

Our vision deliberately preserves the human architect's role in the final, most critical stages of debugging. As others have also stated, the future is in the hands of ``cyborgs and centaurs'', depicting a symbiosis of human ingenuity and artificial intelligence to solve problems together \cite{harvard_cyborgs, centaur_cyborg}. We have consistently found that the most profound insights and innovative ideas for future microarchitectures emerge from the painstaking process of resolving performance discrepancies. When faced with a test where the model is significantly faster than the RTL, the debugging process forces the architect to ask the fundamental question: ``\textit{Why can't the RTL be this fast?}'' This question has frequently been the catalyst for identifying opportunities for subsequent processor generations. By automating the toil but safeguarding the role of human curiosity and analysis, we ensure that PV remains more than just a process of validation; it becomes a virtuous cycle of discovery and innovation.  
\section{Conclusion}

In this paper, we have detailed the pre-silicon performance verification methodology employed in the development of the AmpereOne$^\circledR$ custom CPU core. We have shown that in an era where microarchitectural innovation is paramount, a disciplined PV process is a necessity for delivering a competitive product. Our methodology is founded on the systematic correlation of RTL against a cycle-accurate performance model along with directed microbenchmarks for specific units. We showed our multi-faceted strategy encompassing data-driven workload curation, unified event-stream analysis, and agile high-frequency regressions with triaging.

Our case study of the L2 prefetcher subsystem illustrated a key principle of this work. High-level automated correlation provides the initial signal, but effective root-cause analysis requires targeted debug strategies. The development of the event-replay mechanism to isolate algorithmic logic from system-level timing noise exemplifies the strategies we use to solve complex verification challenges. This blend of broad  automated coverage with deep expert-driven analysis is essential for identifying and resolving bugs that would have otherwise degraded performance in the final silicon.

As we look to the future, the principles of automation and architect-centric tooling remain central. We envision a new generation of agentic AI workflows designed not to replace human experts, but to amplify their capabilities. By automating the laborious tasks of sifting through data and running exhaustive test permutations, these systems will free architects to focus on the creative analysis, as we have found that profound microarchitectural insights are forged in the crucible of performance debug. The process of questioning why a design fails to meet its model's potential helps guide what leads to the next generation of improvements.

Ultimately, this work underscores that performance verification is more than just a final check-off in the design process; it is a cycle of discovery. It is through the rigorous process of reconciling design with intent that a processor is proven to be both functionally correct and capable of delivering on the performance promise that defines its success in the marketplace.

\begingroup
\footnotesize
\bibliographystyle{IEEEtran}
\bibliography{99-citations}
\endgroup

\end{document}